%% file: main.tex
\documentclass[submission, Phys]{SciPost}

\hypersetup{
    colorlinks,
    linkcolor={red!50!black},
    citecolor={blue!50!black},
    urlcolor={blue!80!black}
}

\usepackage[bitstream-charter]{mathdesign}
\DeclareSymbolFont{usualmathcal}{OMS}{cmsy}{m}{n}
\DeclareSymbolFontAlphabet{\mathcal}{usualmathcal}

\usepackage{longtable}
\usepackage{multirow}
\usepackage{caption}
\usepackage{placeins}
\usepackage[labelformat=simple]{subcaption}

\usepackage{enumerate}
\usepackage{xspace}
\usepackage{xcolor}
\usepackage{graphicx}
\usepackage{bookmark}
\usepackage[capitalise]{cleveref}

\usepackage{siunitx}
\usepackage{esvect}
\usepackage{booktabs}
\usepackage{todonotes}

\newcommand{\pthad}{\ensuremath{p_{\mathrm{T}}^{t,had}}}
\newcommand{\pttt}{\ensuremath{p_{\mathrm{T}}^{tt}}}
\newcommand{\mtt}{\ensuremath{m^{tt}}}
\newcommand{\ttbar}{\ensuremath{t \bar{t}}}

\begin{document}

\begin{center}{\Large \textbf{
    Flow Away your Differences:\\ Conditional Normalizing Flows as an
    Improvement to Reweighting}}
\end{center}

\begin{center}
Malte Algren\textsuperscript{1$\dagger$},
Tobias Golling\textsuperscript{1},
Manuel Guth\textsuperscript{1},
Chris Pollard\textsuperscript{2} and
John Andrew Raine\textsuperscript{1$\star$}
\end{center}

\begin{center}
{\bf 1} University of Geneva, Geneva, Switzerland
\\
{\bf 2} University of Warwick, United Kingdom
\\
${}^\dagger$ {\small \sf malte.algren@unige.ch}
${}^\star$ {\small \sf john.raine@unige.ch}
\end{center}

\begin{center}
\today
\end{center}


\section*{Abstract}
{\bf
We present an alternative to reweighting techniques for modifying distributions
to account for a desired change in an underlying conditional distribution, as is
often needed to correct for mis-modelling in a simulated sample.
We employ conditional normalizing flows to learn the full conditional
probability distribution from which we sample new events for conditional values
drawn from the target distribution to produce the desired, altered distribution.
In contrast to common reweighting techniques, this procedure is independent of
binning choice and does not rely on an estimate of the density ratio between
two distributions.

In several toy examples we show that normalizing flows outperform reweighting
approaches to match the distribution of the target. 
We demonstrate that the corrected distribution closes well with the ground
truth, and a statistical uncertainty on the training dataset can be ascertained
with bootstrapping.
In our examples, this leads to a statistical precision up to three times
greater than using reweighting techniques with identical sample sizes for the
source and target distributions.
We also explore an application in the context of high energy particle physics.
}

\vspace{10pt}
\noindent\rule{\textwidth}{1pt}
\tableofcontents\thispagestyle{fancy}
\noindent\rule{\textwidth}{1pt}
\vspace{10pt}

\section{Introduction}
\input{includes/introduction}

\section{Method}
\input{includes/method}
\section{Related work}
\input{includes/related}
\section{Application to a toy example}
\input{includes/results}
\section{Further applications}
\subsection{High energy physics}
\input{includes/results_top}
\section{Conclusion}
\input{includes/conclusion}

\section*{Acknowledgements}
We thank Slava Voloshynovskiy for his helpful discussions and feedback.
MA, MG, and JR are supported through funding from the SNSF Sinergia grant called Robust Deep Density Models for High-Energy Particle Physics and Solar Flare Analysis (RODEM) with funding number CRSII$5\_193716$ and the SNSF project grant 200020\_212127 called "At the two upgrade frontiers: machine learning and the ITk Pixel detector".
MG is further supported with funding acquired through the Feodor Lynen scholarship award.
CP acknowledges support through STFC consolidated grant ref ST/W000571/1.

\bibliography{main}

\clearpage

\begin{appendix}
\input{includes/appendix}

    
\end{appendix}

    
        

\nolinenumbers

\end{document}

%% file: includes/introduction.tex

In many areas of science, simulations are used to depict the behaviors of real world systems.
However, due to lack of knowledge or approximations applied to make the
simulation tractable, predicted distributions often deviate from observations at
some level of precision. 
For accurate tests of theoretical predictions using data collected from
experiments, it is therefore necessary to correct the simulated data and adjust
the distributions to better match observed data.
Correcting observed mis-modelling is particularly important in high energy
particle physics~(HEP), in which Monte Carlo simulation~(MC) of complex
processes form the bedrock of tests of the Standard Model of particle
physics~(SM); for example see Refs.~\cite{Buckley:2011ms, ATLAS:2010arf,
GEANT4:2002zbu}.

One way of improving a multi-dimensional density $p$ is to alter the initial
marginal densities of some quantities $p(c)$ to better reflect a target $q(c)$;
$q(c)$ may represent an observation or ground truth of some control variable,
for example.
Quantities of interest, $x$, will have simulated and true conditional densities
on $c$, respectively $p(x|c)$ and $q(x|c)$.
A new joint density can be constructed from the conditional density of $p$
together with the marginal density over $c$ of $q$: $p^\prime(c, x) = p(x|c)
q(c)$.
Following the nomenclature of Cover and Thomas, the relative
entropy between any two joint densities $f$ and $g$ is given by
\begin{equation*}
  D(f(c, x) \ ||\  g(c, x)) = D(f(c) \ ||\  g(c)) + D(f(x|c) \ ||\  g(x|c)),
\end{equation*}
proven as Theorem~2.5.3 in Ref.~\cite{cover2012elements}.
From this is it clear that
\begin{equation}
  \label{eq:DKLprime}
  D(p^\prime(c, x) \ ||\  q(c, x)) \le D(p(c, x) \ ||\  q(c, x)),
\end{equation}
since $\forall f, g.\ D(f \ ||\  g) \ge 0$, while
$D(p^\prime(c) = q(c) \ ||\  q(c)) = 0$.
In other words, the divergence between $p^\prime(c, x)$ and $q(c, x)$ comes
entirely from any residual difference in their conditional densities, $D(p(x|c)
\ ||\  q(x|c))$.


The standard approach in HEP to correct or alter a distribution is to derive the
density ratio estimates (DREs) between the initial and target distributions in
some space.
The density ratio at a point $c$ is defined as $q(c) / p(c)$, where $q$ and $p$
are the two densities being compared.
These DREs can be applied as weights to samples of the predicted density $p$ to
improve the matching of the simulated distribution to the some other
distribution $q$ -- the observed data, for example.
Alternatively, the DREs can be used to down-sample the initial distribution to
match the target.
Histograms or classifiers are commonly used to estimate the density ratios and
derive weights for each event.

In this work we explore the use of conditional Normalizing
Flows (cNFs)~\cite{tabak_flows,dinh2014nice,razende_flows,dinh2016density,flows_review,CNFs}
to adjust the distributions of samples drawn from a simulator.
We demonstrate that using cNFs, we can achieve better closure than both binned
and unbinned reweighting approaches on a non-trivial toy example, and
demonstrate its further possible applications to a HEP example, correcting the
kinematics of top quark pairs produced at the LHC.

%% file: includes/method.tex
\subsection{Reweighting}

In HEP, the standard approach to match distributions over $c$ and propagate to
other distributions over quantities of interest $x$ is to derive the DREs as a 
function of $c$ ($R(c) = q(c) / p(c)$ for an initial distribution $p$ and a
target distribution $q$).
The values $R(c)$ are applied to samples drawn from the initial distribution
$c,x\sim p(c,x)$ as weights, such that the relative entropy between the altered
probability distribution matches $p^\prime(c,x) = p(x | c)q(c)$.
As shown in Equation~\ref{eq:DKLprime}, $p^\prime(c,x)$ is guaranteed to have a 
lower relative entropy with respect to the target distribution $q(c,x)$ than is
observed between the initial distribution $p(c,x)$ and $q(c,x)$.
Given the target distribution $q(c,x)$ has values $c\sim q(c)$,
it follows that with weights calculated as 
\begin{equation*}
    R(c)=\frac{q(c)}{p(c)},
\end{equation*}
and starting from the definition of the joint distributions
\begin{equation}
    p_{c\sim p(c)}(c,x)=p(x|c)p(c),
    \label{eq:jointprob}
\end{equation}
it can straightforwardly be shown that
\begin{align*}
    p(c,x) R(c) &= p(x|c)p(c)R(c),\\
    p(c,x) R(c) &= p(x|c)p(c)\frac{q(c)}{p(c)},\\
    p(c,x) R(c) &= p^\prime(c,x).
\end{align*}
In deriving the weights $R(c)$, one needs access only to the marginal
distributions $p(c)$ and $q(c)$, which can for instance be calculated from the
samples drawn from the two distributions.

Two methods are commonly used used to extract $R(c)$ in HEP.
Perhaps the simplest is to build the ratio from two binned histograms, with
the benefit that they are straightforward to calculate.
Problems arise when the true $R(c)$ changes on a scale smaller than the width of
the histogram bins; in large-dimensional spaces with a finite number of samples
from the initial and target distributions, one is often forced to choose between
small bin sizes with an imprecise DRE estimate in each or large bins over which
$R(c)$ may change substantially.
This may be mitigated somewhat through sequential application of weights
calculated in each dimension separately, but this procedure correctly produces
the target distribution only in the limit that the joint densities factorize
into a product of marginals.

A second approach is to approximate $R(c)$ using neural networks.
As presented in Ref.~\cite{CARLCranmer}, it is possible to train a classifier
$f_\phi(c)$, by minimizing over the classifier parameters $\phi$, to
discriminate between samples drawn from the distributions $p(c)$ 
and $q(c)$, from which the resulting classifier output can be converted
into a weight with 
\begin{equation}
    R(c) = \frac{f_\phi(c)}{1-f_\phi(c)}.
    \label{eq:dreweights}
\end{equation}
This has advantages over the binned approach in that it is fully continuous and
can correlate weights at the required length scales in order to build the
distribution $p^\prime(c,x)$.

However, both approaches based on DREs share the same drawback: they
rely on building an approximate ratio of two probability distributions rather
than simply approximating $p(x | c)$ and altering the density over $c$.
If there is little overlap between the support of the two distributions this
leads to either very large or very small values of $R(c)$, resulting in an
effective loss of the sample size once reweighting is applied.
This can have a large impact on the statistical precision of $q(c,x)$.

\subsection{Conditional normalizing flows}

Normalizing flows
are a family of generative models which in recent years have gained traction due
to their ability to map the probability distribution of a complex distribution
from a simpler distribution of the same dimensionality.
They learn the density of the complex distribution under a series of invertible
transformations $f_\theta$, a family of functions parameterized by $\theta$, by
minimizing the loss function
\begin{align*}
    \label{eq:flowloss}    
    \log\left(p_X(x)\right) &= \log\left(p_Z(f_\theta^{-1}(x))\right) - \log\left|\det\left(\mathcal{J}(f_\theta^{-1}(x))\right)\right|
\end{align*}
with respect to $\theta$, where $\mathcal{J}\left(f_\theta^{-1}(x)\right)$ is
the Jacobian of the transformation $f_\theta^{-1}(x)$, which maps the complex
distribution $p_X$ to the chosen base distribution $p_Z$.
Conditional normalizing flows similarly follow from the change of variables
formula, with the probabilities and transformation conditional on some
additional properties $c$
\begin{equation}
    \label{eq:cflowloss}
    \log\left(p_X(x|c)\right) = \log\left(p_Z(f_\theta^{-1}(x|c))\right) - \log\left|\det\left(\mathcal{J}(f_\theta^{-1}(x|c))\right)\right|.
\end{equation}

Due to their ability to learn the conditional probability distribution $p(x|c)$
from samples drawn from $p(c,x)$, we demonstrate normalizing flows can be used
directly with Eq.~\ref{eq:jointprob} to perform the same role as reweighting. 
Instead of learning or approximating the density ratio, one needs only to sample
from the normalizing flow but with the target distribution $q(c)$ to produce the
desired distribution $q(c,x)$.

There are several advantages to this approach over constructing DREs and
reweighting.
The statistical precision of regions of probability space are no longer
constrained by the derived DREs, as all sampled events have a weight of unity.
Instead, the statistical uncertainty comes only from the statistical precision
of the learned conditional density $p(x|c)$, as it is possible to sample any
number of events from it for each $c$.
Where necessary, the statistical uncertainty on the sampled distribution can be
estimated through bootstrapping~\cite{Efron:526679}.

In our tests, the normalizing flows are able to learn a more precise approximation of the
conditional probability distribution $p(x|c)$ at low probability values for $c$
than is possible when considering only data with those values $c$.
This is a result of the normalizing flow learning the correlations in the underlying
probability distribution function.
Similar behaviour is observed when training neural networks to approximating the DRE
in comparison to binned approaches.

%% file: includes/related.tex

Modifying distributions to account for mis-modelling or to follow some other
underlying distribution is often performed with event weights or resampling.

In the case of reweighting, machine learning approaches based on density ratio estimation have had a large amount of success, in particular in the field of high energy physics.
The CARL method, introduced in Ref.~\cite{CARLCranmer}, uses neural networks to learn the likelihood ratio which can be used for covariate shift and importance sampling of distributions.
In Ref.~\cite{Rogozhnikov:2016bdp}, boosted decision trees are employed to derive event weights for reweighting.
In Ref.~\cite{Andreassen:2019nnm}, neural networks are used to reweight the full phase space of events generated with two different MC generators, and extended to the conditional case in Ref.~\cite{Nachman:2021opi}.

Instead of reweighting events, it is also possible to learn a mapping between the initial and target distributions.
In Ref.~\cite{Pollard:2021fqv} partially input convex neural networks~\cite{icnn} are used to calibrate distributions from MC simulation to match those observed in data.
In Refs.~\cite{Choi:2020bnf,Golling:2022nkl} normalizing flows are trained to transport events from MC to data domain to another, and in Ref.~\cite{Raine:2022hht} they are used to move to data to different conditional values on the same distribution.

In our work we use normalizing flows to learn the conditional probability distribution of the initial data, from which events following a different conditional distribution can be sampled.
This is similar to the approach used in Ref.~\cite{Hallin:2021wme}.
Here, a normalizing flow is trained to learn the conditional density on side-band data and apply it to a blinded signal region.
These data are used to train a classifier, and no treatment is required or derived to account for the statistical uncertainty on the relating distribution.
Concurrent with our work, Ref.~\cite{Hartman:2022clj} presents an examination of similar methods applied for estimating backgrounds in blinded signal regions, including estimation of the statistical uncertainty.

%% file: includes/results.tex
To show the benefit of cNFs in comparison to traditional reweighting approaches,
we use a toy example from which any number of data can be sampled for any chosen
change of conditions. 
As shown in Ref.~\cite{Butter2020qhk}, sampling more data points from a
generative model than are available in the initial dataset can lead to higher
statistical precision in the output of trained models.

We define a two dimensional probability distribution $f(c)$
solely dependent on two conditional variables $c = (c_0, c_1)$, 
with marginals $f_i$, $i\in 0,1$
\begin{align}
    f_0(c'_0, c'_1) = \frac{c'_0}{(1+c_1^{'2})}-\frac{c'_1}{(1+c_0^{'2})} \qquad \textrm{with} 
    \qquad c'_0, c'_1 = \mathcal{N}\Biggl(\begin{pmatrix} c_0 \\ c_1 \end{pmatrix} ,
    \begin{pmatrix} 2 \cdot c_0,c_1 \\
        c_0,c_1/2 \end{pmatrix} \Biggr) \\
    f_1(c'_0, c'_1) = \frac{(c'_0+c'_1)^2}{(1+c'_0+c'_1)}-\frac{(c'_0-c'_1)}{(1+c_0^{'2})} \qquad \textrm{with} 
    \qquad c'_0, c'_1 = \mathcal{N} \Biggl( \begin{pmatrix} c_0 \\ c_1 \end{pmatrix} ,
    \begin{pmatrix} c_0,c_1/2 \\ c_0,c_1/2 \end{pmatrix} \Biggr)
    \label{eq:toy_functions}
\end{align}
from which samples $x$ are drawn.
Here $c_0^\prime$ and $c_1^\prime$ are distributions defined by $c_0$ and $c_1$.

\begin{figure}[htpb]
    \includegraphics[width=\textwidth]{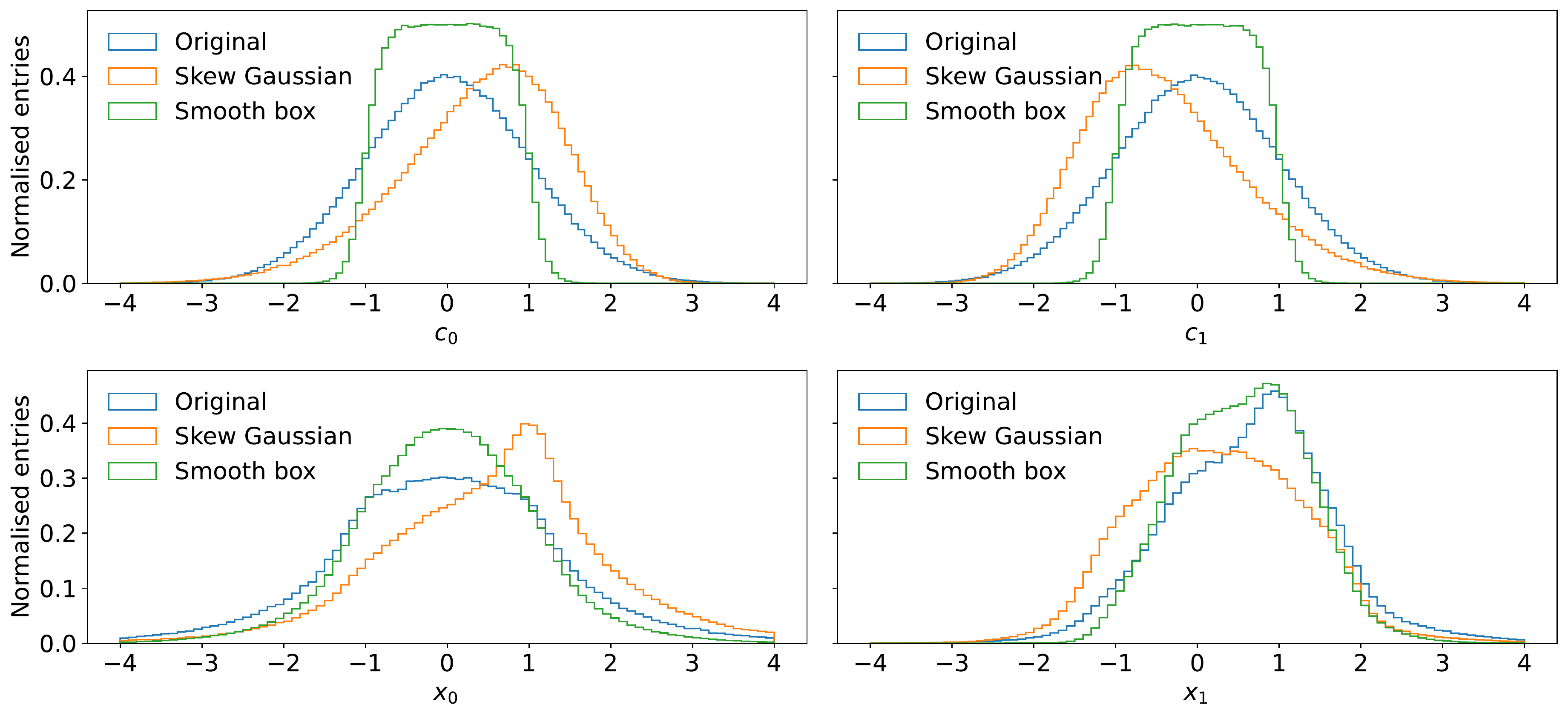}
    \centering
    \caption{
        The ground truth distributions of the conditional quantities $c_0$ and
        $c_1$ (upper panels) as well as the marginals of the quantities of
        interest, $f_0$ and $f_1$ (lower panels), for three choices of
        distribution over $c_0$ and $c_1$: a standard multivariate normal
        Gaussian distribution, a skewed Gaussian, and a uniform distribution
        over $[-1, 1]$ for each choice of $c_i$.
    }
    \label{fig:toydists}
\end{figure}

For the initial probability distribution, we choose a 2D standard normal distribution for $p(c)$,
$c\sim\mathcal{N}\left(0,\mathbb{1}\right)$.
Figure~\ref{fig:toydists} shows the 1D marginals of $x_0$ and $x_1$ for samples drawn from $f(c)$. 
For target distributions we change the distribution from which $c$ is
sampled.
Here we use a skewed Gaussian distribution, with
\begin{equation*}
    c_1\sim\mathcal{S}\mathcal{N}\left(1.5,1.5, -2.5\right) \,\qquad \textrm{and} 
    \qquad
    c_2\sim\mathcal{S}\mathcal{N}\left(1.5,1.5, 2.5\right), 
\end{equation*}
where the three parameters to $\mathcal{S}\mathcal{N}$ are the location, scale,
and shape, respectively.
We also test a symmetric smooth box distribution defined by
\begin{equation*}
    c_{1/2}\sim q(x) = \frac{1}{1 + \exp\bigl(-10(x + 1)\bigr)} - \frac{1}{1 + \exp\bigl(-10(x - 1)\bigr)}.
\end{equation*}
Each target distribution has a sample size of $10^7$, such that the statistical
uncertainty on the target distribution does not impact the performance of the
three methods.

We train a conditional normalizing flow on the initial samples, learning the
conditional density $p(x | c)$ from samples $x \sim f(c)$.
To produce the target distributions with different conditions, we sample once
from the cNF for all $10^7$ values of $c$ from the target conditions to
generate the target distribution under new conditions.

We compare the cNF approach to two reweighting references.
For \emph{DRE (binned)} we take a ratio of the initial and target
distributions over $c$ in two dimensions, and for \emph{DRE (NN)} we
train a classifier following the procedure in Ref.~\cite{CARLCranmer} with $c$
as an input.
The distributions of $f(c)$ for the three scenarios are shown in
Fig.~\ref{fig:toydists}. 

\subsection{Configuration}

We construct our conditional normalizing flows using 5 stacks of rational
quadratic splines~\cite{NeuralSplines} with autoregressive transformations for
each layer.
The cNFs are implemented using the \texttt{nflows}~\cite{nflows} package with
PyTorch~\cite{Pytorch}. 
The networks are trained for 500 epochs with a cosine annealing learning
rate~\cite{DBLP:journals/corr/LoshchilovH16a}, with an initial value of
$10^{-4}$ using the \texttt{Adam} optimiser~\cite{DBLP:journals/corr/KingmaB14}
and a batch size of 512.

For both of the reweighting approaches, event weights are calculated for each
target distribution.
For DRE (binned), the weights are calculated by taking the ratio of the
target and initial histograms, defined using a uniform grid with 100 bins
between [-4,4] on each axis in $c$.
For DRE (NN), the classifier trained on $c$ from the target and initial
distribution comprises four layers of 64 nodes, with \texttt{ReLU} activations
in the hidden layers and a sigmoid activation on the output.
It is trained with a flat learning rate of $10^{-4}$ for 50 epochs using the
\texttt{Adam} optimiser.
Event weights are defined using Eq.~\ref{eq:dreweights} with the classifier output.

\subsection{Comparison of methods}

\begin{figure}[htpb]
    \centering
    \subfloat[\centering Skewed Gaussian]
    {\includegraphics[width=\textwidth]
        {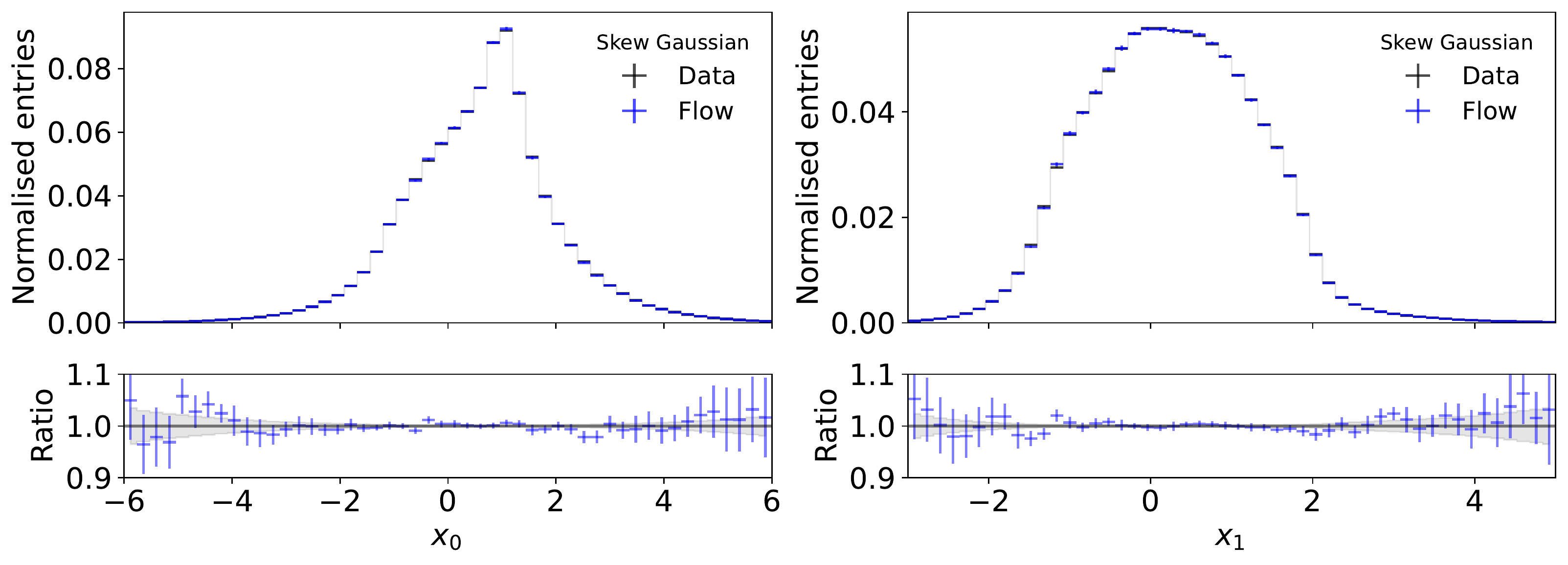}
    } 
    \\
    \vspace{20pt}
    \subfloat[\centering Uniform]
    {\includegraphics[width=\textwidth]
        {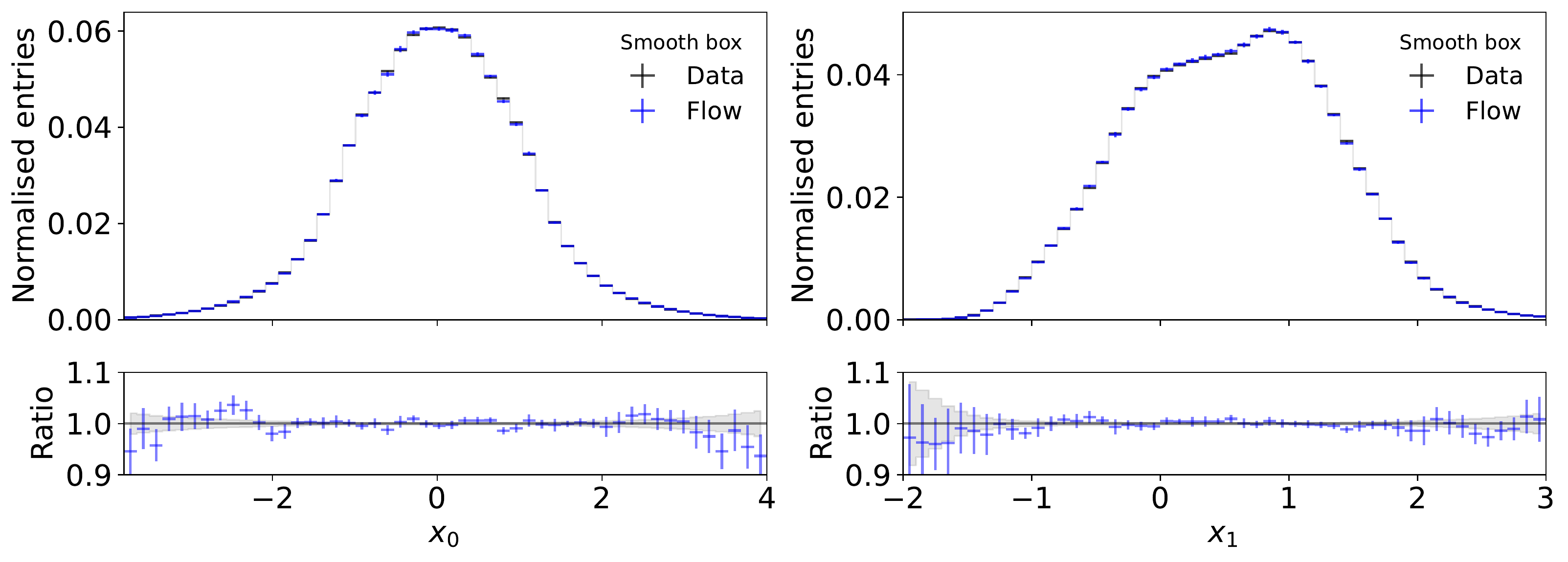}
    } 
    \caption{
        The marginals of the cNF trained on the original Gaussian conditional
        distributions applied to a skewed Gaussian or uniform conditional
        distribution compared to the ground truth.
        Good closure is observed for both marginals.
        Statistical uncertainties in the cNF function are estimated using
        12 bootstraps.
    } 
    \label{fig:toy_result1}
\end{figure}

\begin{figure}[htpb]
    \centering
    \subfloat[\centering DRE (binned)]{{\includegraphics[width=\textwidth]{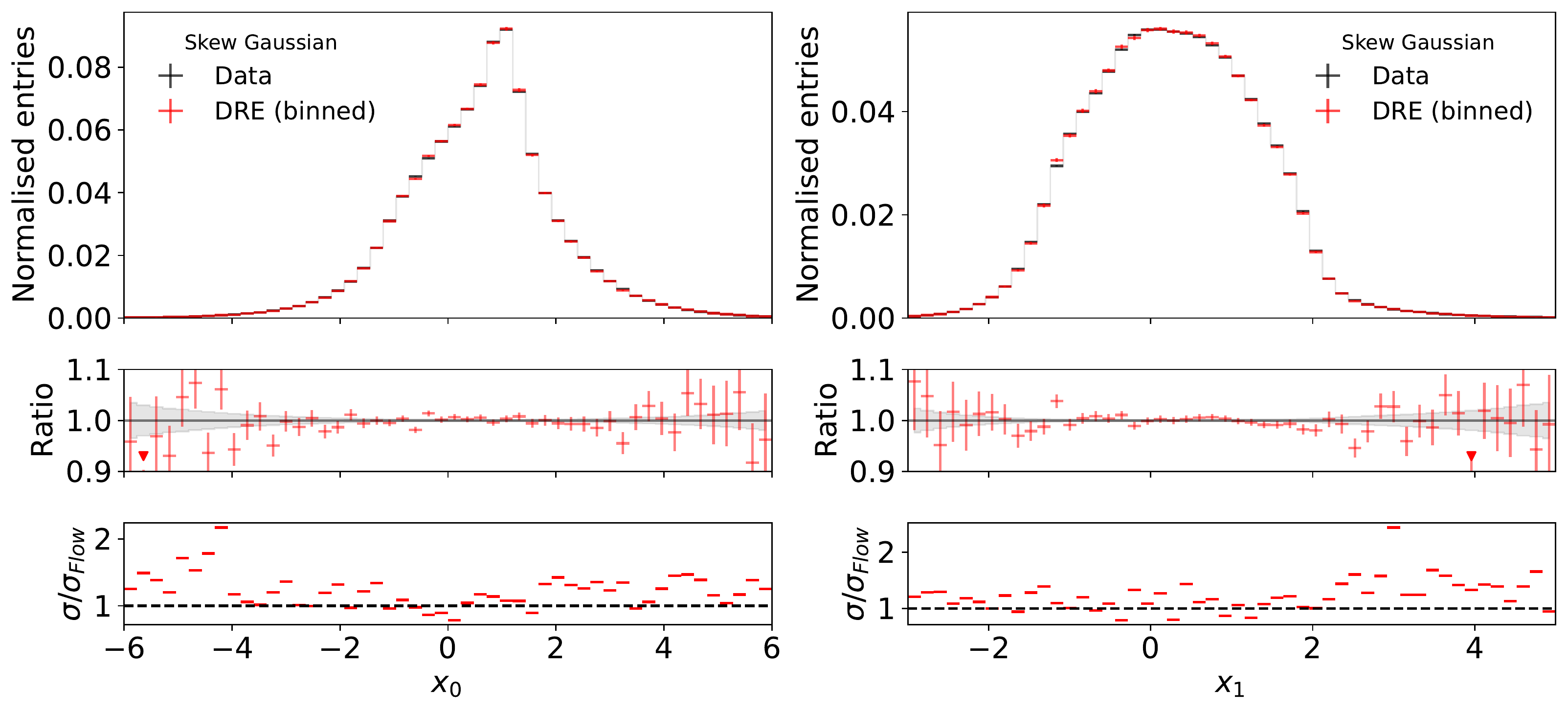} }}%
    \\
    \vspace{20pt}
    \subfloat[\centering DRE (NN)]{{\includegraphics[width=\textwidth]{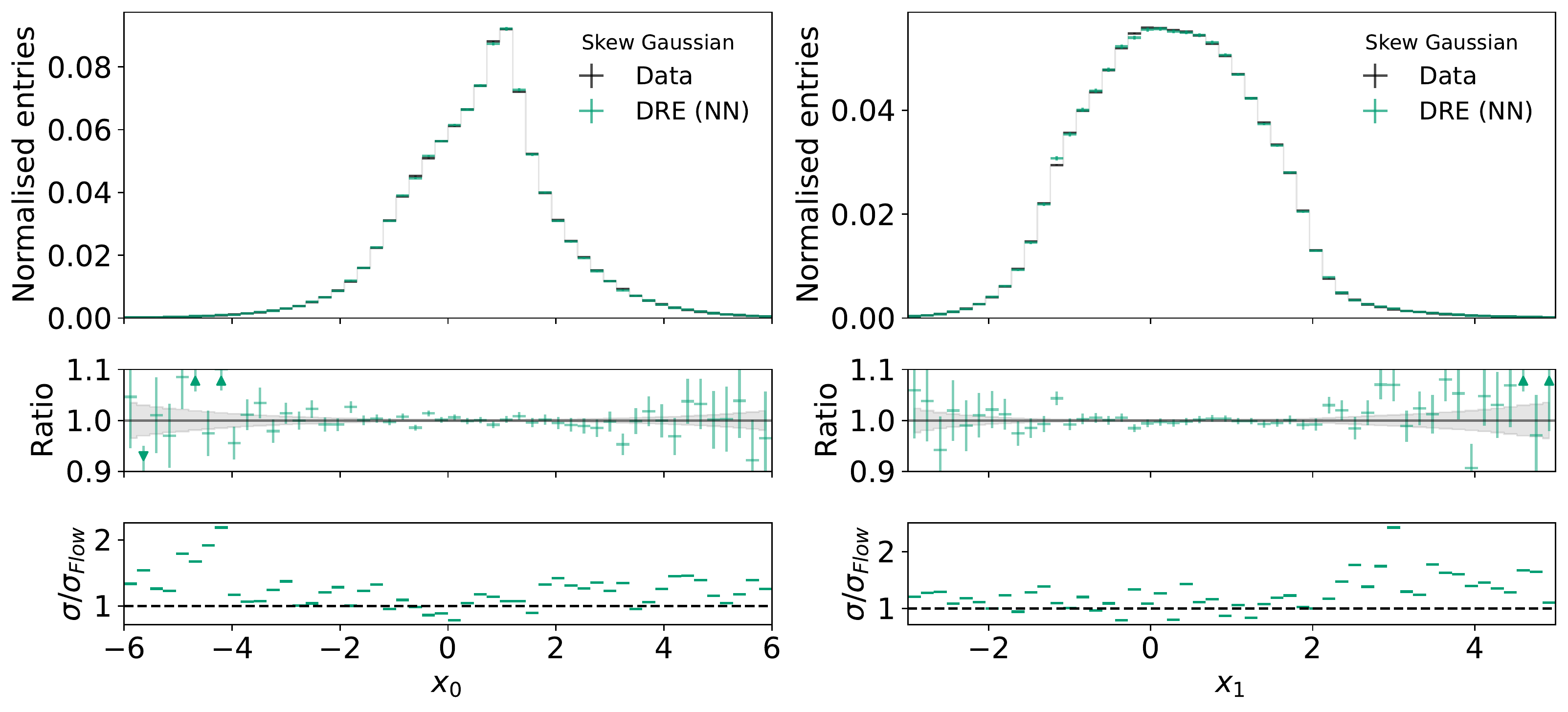} }}%
    \caption{
    Closure of corrected conditional distributions to the ground truth when
    using the reweighting methods with a skewed Gaussian distribution over the
    conditional quantities.
    The marginal distributions of $f_0$ and $f_i$ is shown in the upper panels,
    the ratio to the ground truth distribution in the middle panels, and the
    ratio between the statistical uncertainty on the reweight methods and that
    from the conditional normalizing flows in the lower panel.
    The conditional normalizing flows generally result in strongly reduced
    statistical uncertainties compared to the reweight methods.
    }
    \label{fig:marginal_plots_of_other_methods_skew}
\end{figure}

\begin{figure}[htpb]
    \centering
    \subfloat[\centering DRE (binned)]{{\includegraphics[width=\textwidth]{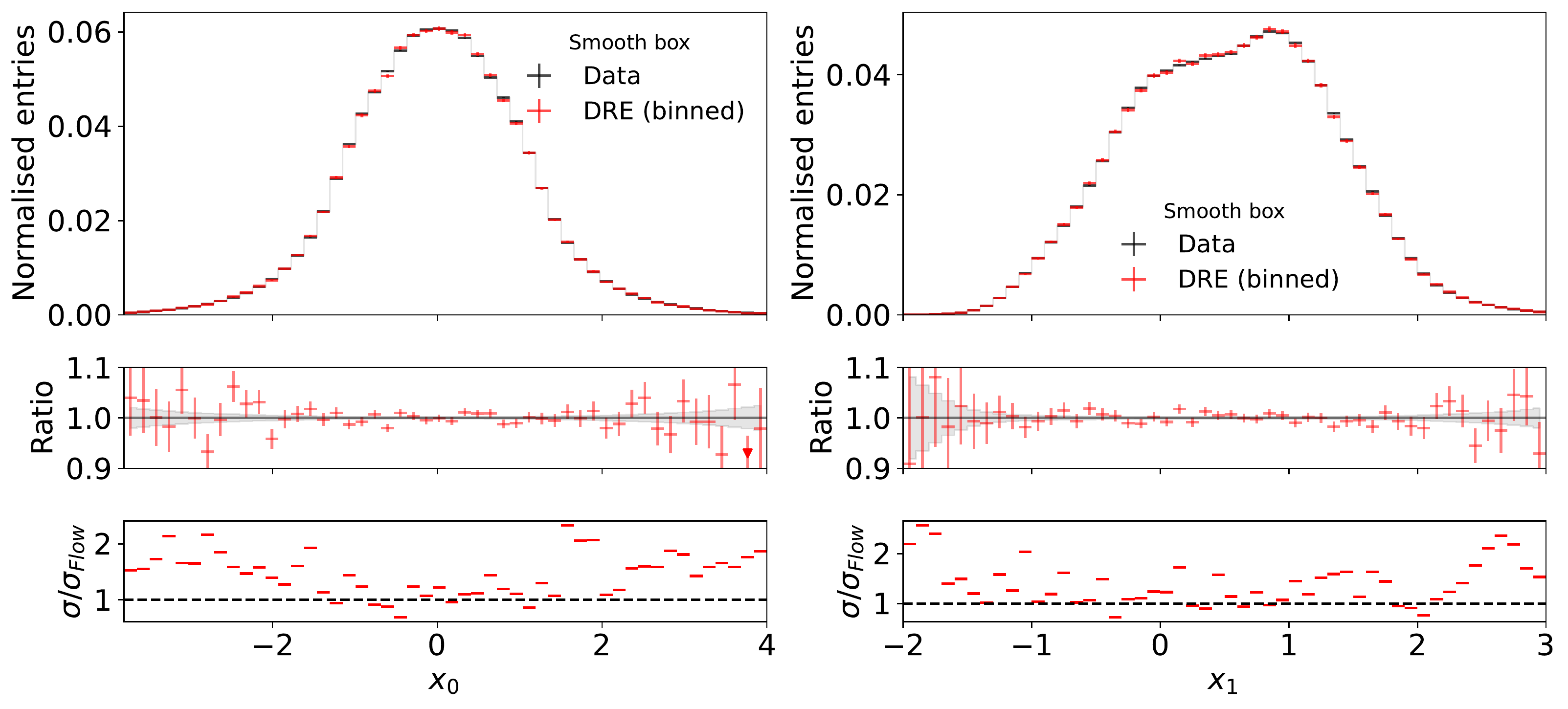} }}
    \\
    \vspace{20pt}
    \subfloat[\centering DRE (NN)]{{\includegraphics[width=\textwidth]{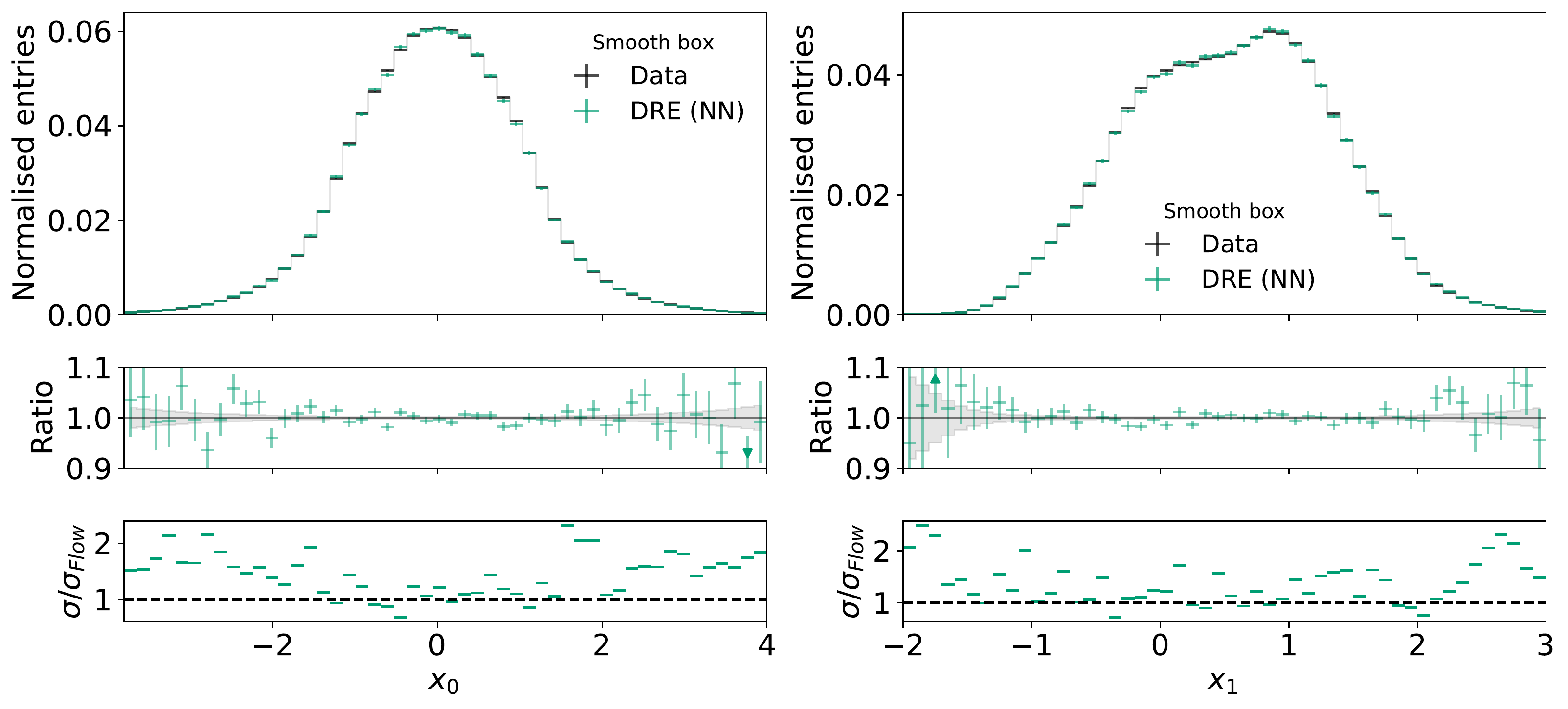} }}
    \caption{
    Closure of corrected conditional distributions to the ground truth when
    using the reweighting methods with a uniform distribution over the
    conditional quantities.
    The marginal distributions of $f_0$ and $f_i$ are shown in the upper panels,
    the ratio to the ground truth distribution in the middle panels, and the
    ratio between the statistical uncertainty on the reweight methods and that
    from the conditional normalizing flows in the lower panel.
    The conditional normalizing flows generally result in strongly reduced
    statistical uncertainties compared to the reweight methods.
    }
    \label{fig:marginal_plots_of_other_methods_uniform}
\end{figure}

In Fig.~\ref{fig:toy_result1} we show the distribution of data drawn from the cNF trained on the initial data, but following the Skew Gaussian and Smooth box distributions.
The statistical uncertainty is obtained with bootstrapping, trained 12 cNFs on bootstrapped training data.
The central estimation is taken as the mean over all 12 models.
We compare this to the closure obtained by the DRE (binned) and DRE (NN) approaches shown in \cref{fig:marginal_plots_of_other_methods_skew,fig:marginal_plots_of_other_methods_uniform} for the skewed Gaussian and smooth box target conditional distributions respectively.
In these figures, we look at not only the closure between the reweighting method and the target, but also the relative statistical precision in each bin with respect to the cNF.
The statistical uncertainty on the reweighting approaches is calculated from the sum of weights in each bin.

For both target distributions, the agreement of all approaches is reasonable, however we observe that the cNF is much closer to the true target distribution than the reweighting approaches.
This is most noticeable in the tails of the distribution.
In regions of the distribution with low numbers of events, we observe fluctuations in all three approaches about the true target distribution.
In regions of high yields, we see that the cNF is matching the true target distribution perfectly whereas the reweighting approaches, although still within the uncertainties of the prediction, show less precise closure.
Furthermore, the statistical precision when using the cNF is almost always greater than either reweighting approaches.

As a further measure of performance we train additional classifiers on the output
of each of the three approaches to separate them from the target distribution.
In the case where the distributions match perfectly the classifier will not be
able to separate the samples, whereas the worse the closure the easier it will
be.
All methods result in a much better closure than the initial distribution, with area under the receiver operator
characteristic curve close to 0.5 in all three cases.

\begin{figure}[htpb]
    \centering
    \includegraphics[width=1 \textwidth]{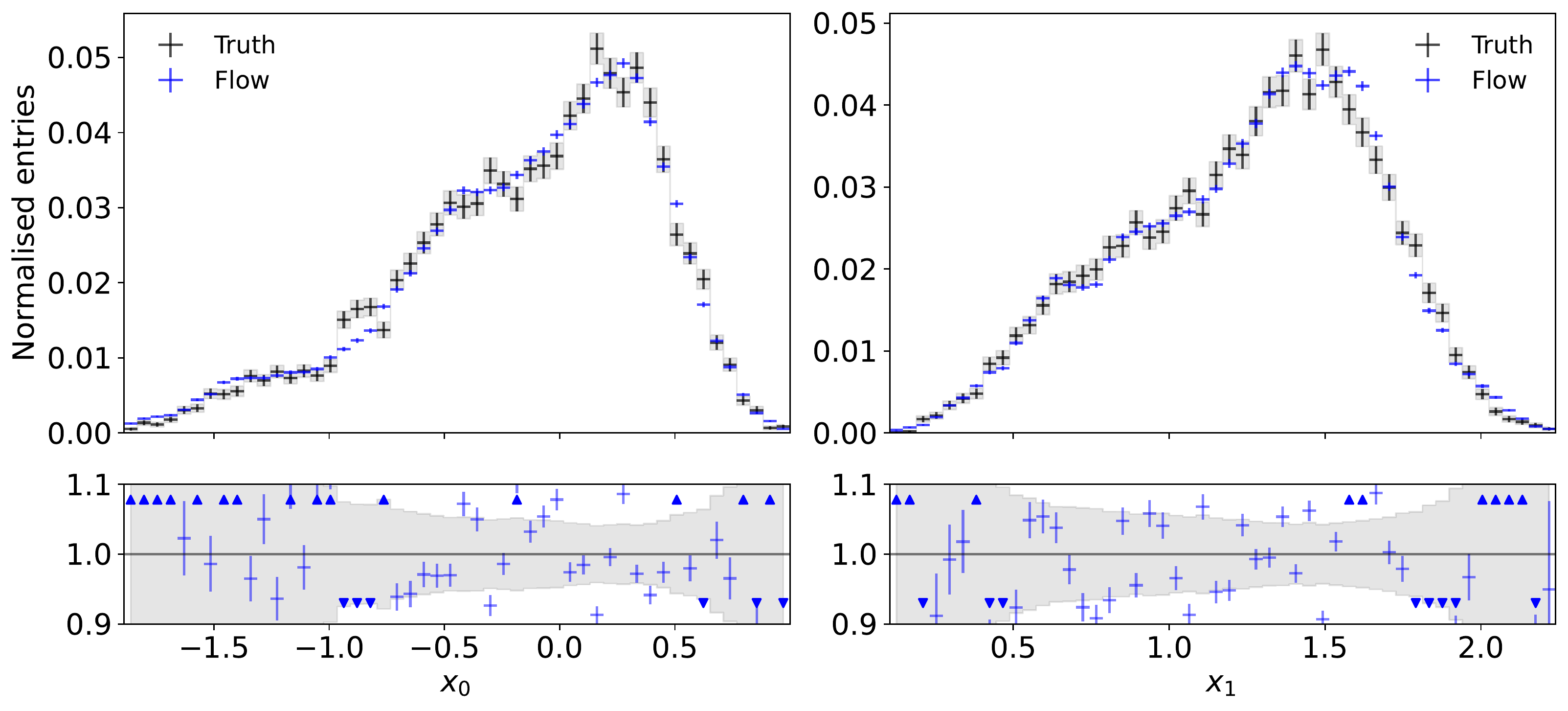}
    \caption{Closure in a small window of conditional values $c_0, c_1 \in (0.9, 1.1]$, for $c_0,c_1 \sim \mathcal{N}(0,1)$.
    Samples generated with the normalizing flow (blue) are compared to data drawn from the original distribution (black).}
    \label{fig:small_conds_window}
\end{figure}

As an additional test we want to verify whether the normalizing flow truly 
learns the conditional probability density from the training samples.
In \cref{fig:small_conds_window} we draw samples from a narrow window in $c$ following the original distribution
over conditions used to train the normalizing flow.
These data are compared to samples drawn from the initial distribution for the same values $c$.
Here we can see that samples from the normalizing flow closely match the original target distribution within statistical uncertainties.

\subsection{Validating statistical uncertainties}

\begin{figure}[htpb]
    \centering
    \subfloat[\centering Skew Gaussian]{{\includegraphics[width=0.45 \textwidth]{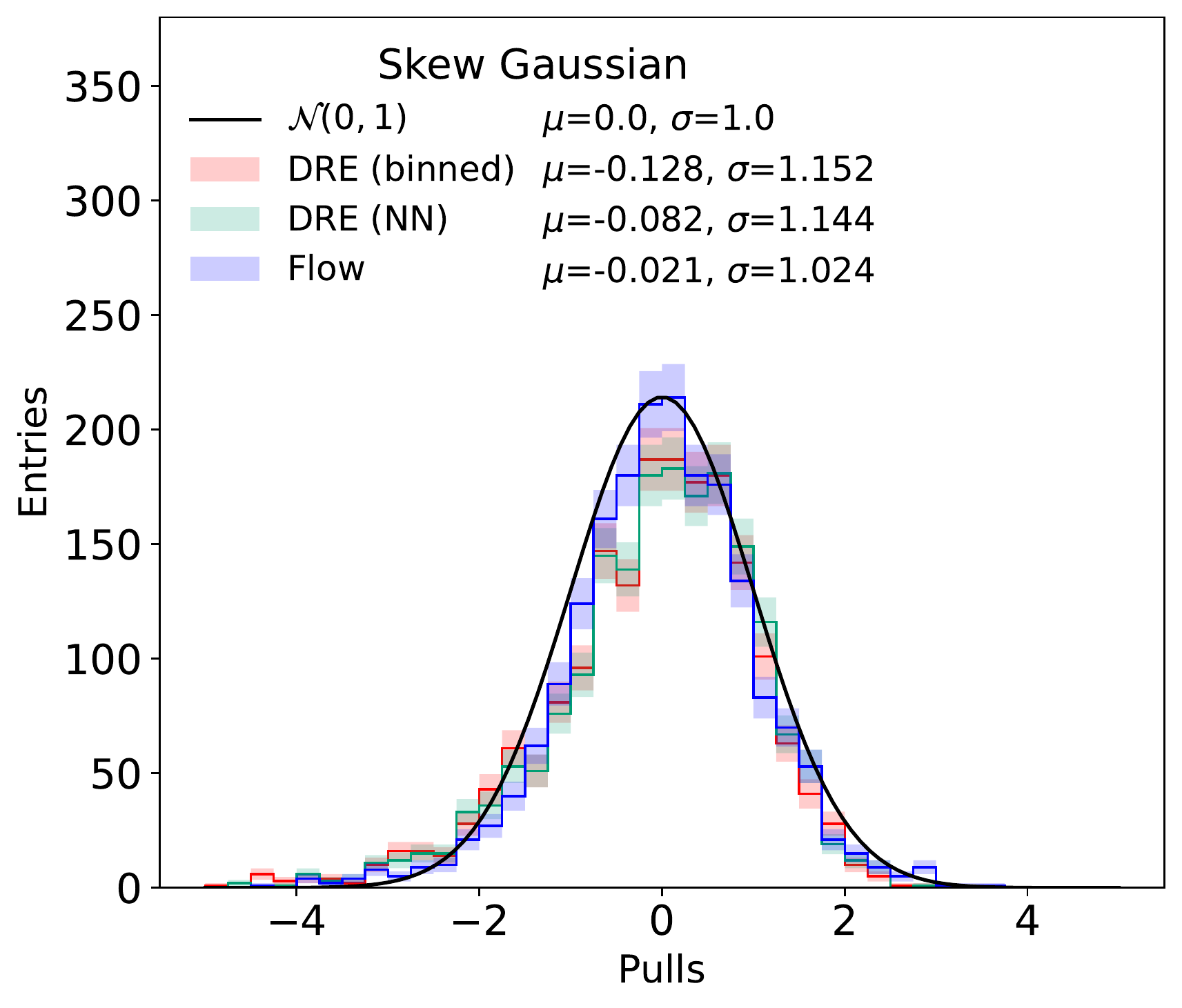} }}%
    \subfloat[\centering Smooth box]{{\includegraphics[width=0.45 \textwidth]{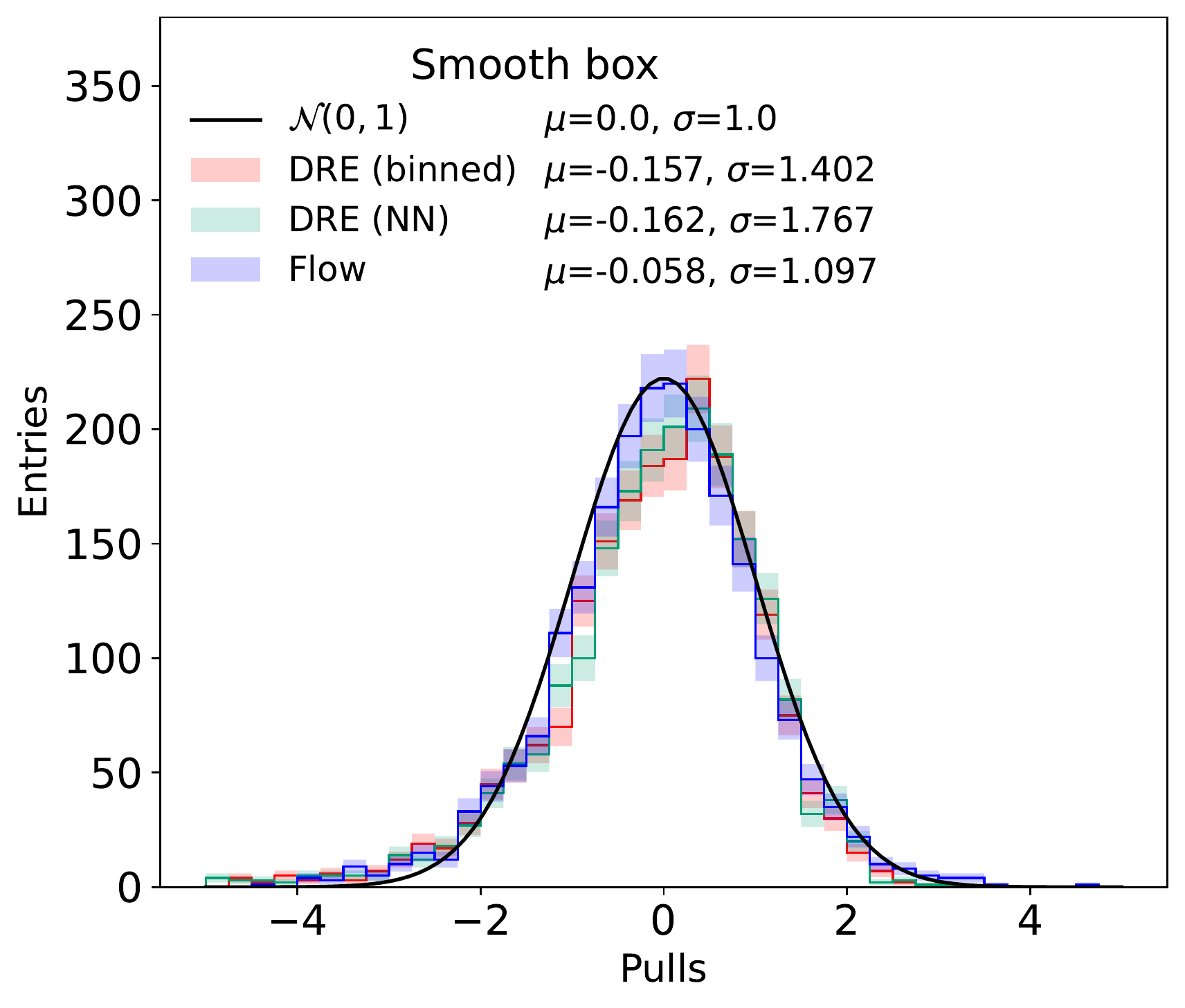} }}%
    \caption{
    Comparisons are shown of pull distributions over the 2D probability density
    $p(f_0, f_1)$, approximated using 2D histograms, for the various strategies
    for correcting distributions over the conditional quantities.
    The normal distribution is shown for reference in black.
    Despite resulting in substantially smaller statistical uncertainties on the
    corrected $p(f_0, f_1)$ prediction, the cNFs maintain appropriate behaviour
    in the pull distribution, indicating a correct estimate of the statistical
    uncertainty is obtained via bootstraps.
    } 
    \label{fig:toy_bootstrap_pulls}
\end{figure}

To verify that the statistical uncertainties associated with the methods are reasonable, we look at the pulls observed in all bins of the two-dimensional distributions.
The pull per bin $i$ is calculated from the observed deviation from the target
\begin{equation}
    \mathrm{Pull}_i = \frac{x_i^{\mathrm{pred}} - x_i^{\mathrm{target}}}{\sigma_i},
\end{equation}
given the total statistical uncertainty $\sigma_i$ of the predicted and target data.
The distribution over all pulls for the distribution should follow a normal distribution.
The pull distributions are compared for the three approaches in \cref{fig:toy_bootstrap_pulls}, where in both cases all three follow a normal distribution.
This demonstrates that not only does the cNF result in higher statistical precision, but also that the uncertainties from bootstrapping correctly estimate the true statistical precision.
In the case of both reweighting approaches, the statistical uncertainties are derived only from the values of the weights, but the uncertainty on the weights themselves is not included.


Additionally, in practice the target distribution over the conditionals, $q(c)$, is not always analytically known.
The limited statistics of target values therefore introduce
an additional source of statistical uncertainty into the reweighting approaches.
For the cNF this would not change the precision of $p(x|c)$, and the impact on the generated samples for the target distribution would be reduced by sampling multiple times for each $c$ from the base density of the normalizing flow.

One area in which all three approaches lose precision is the case where the target distribution $q(c)$
does not fully overlap with $p(c)$ ($\forall c^\prime\sim q(c)$ where $p(c^\prime)=0$).
For these cases the density ratios $R(c)$ are infinite, however as there are no corresponding
data in $p(x,c)$ to apply the weights to, the probability density $p^\prime(x,c)=0$.
For the conditional flow $p^\prime(x,c)$ can have non zero probability density for these target values $c$,
however this depends strongly on the transformations used and whether it is an interpolation or extrapolation region~\cite{158898}.
In the case of interpolation, many neural network architectures are well behaved and the resulting
distributions can be expected to be covered by the statistical uncertainty obtained through bootstrapping.
However, for extrapolated values of $c$ the exact behaviour depends strongly on the transformations used in
the normalizing flow, and on any regularisation used.
It is not expected that the flow will perform well far away from the training region.

%% file: includes/results_top.tex


In many cases, disagreements are observed between the data collected by
experiments and predictions; these predictions are often produced via MC
sampling generators, such as Herwig~\cite{Herwig}, Pythia~\cite{Pythia}, and
Sherpa~\cite{Sherpa}.
For example, there is a long-standing discrepancy between various MC generator
predictions and LHC collision data in the differential cross sections of pairs
of top-quarks as a function of their momenta transverse to the incoming proton
beams ($p_T$).
Top-quark pair production is ubiquitous in analyses of LHC data due to its
relatively high cross-section and striking experimental signature; as such, it
has been the subject of intense scrutiny at the LHC.
Both the ATLAS and CMS Collaborations have produced detailed measurements of
this process and have compared these measurement to state-of-the-art
predictions~\cite{CMS:2012hkm,ATLAS:2015dbj,ATLAS:2015lsn,CMS:2015rld,ATLAS:2019hxz,CMS:2021vhb}.
Both experiments have reported differences between data and MC generator
simulations.
Analyses of the LHC data for which top-quark pair production is a large
background often resort to a binned reweighting strategy to account for this
discrepancy (see, for example, Refs.~\cite{ATLAS:2020lks, ATLAS:2018rvc}).
By doing so, they hope to mitigate any mis-modeling of this background process
and it impact on their primary analysis target.

In this example we use apply conditional normalizing flow approach to resample
$t\bar{t}$ events from simulation according to the measured distributions from
data in one kinematic distribution and observe the impact on other measured
kinematic distributions.
The measured kinematic observables are described in Table~\ref{tab:ttbar_vars},
and are chosen following a recent measurement of differential $t\bar{t}$ production cross
section performed by the ATLAS collaboration~\cite{ATLAS:2017cez}.

\begin{table}[h]
    \caption{Kinematic observables at the particle level for events in top quark pair production in the single lepton channel.}
    \label{tab:ttbar_vars}
    \centering
    \begin{tabular}{c | l}
        \toprule
        Observable & Description\\
        \midrule
        \pthad & Transverse momentum of the hadronically decaying top quark\\
        \mtt & Invariant mass of the top quark pair\\
        \pttt & Transverse momentum of the top quark pair\\
        \bottomrule
    \end{tabular}
\end{table}

For this study we choose to correct the modelling of the 
transverse momentum of the top quark \pthad\ to match that in
data.
However, in principle the method can be used for any combination of the
chosen observables, with the impact observed on the remaining distributions.

Top-quark pair production cross sections were predicted with
Pythia~v8.3~\cite{Pythia} using the Monash tuned set of
parameters~\cite{Skands:2014pea} to simulate $t\bar{t}$ production at the LHC,
corresponding to leading order (LO) + parton shower (PS) accuracy.
The NNPDF2.3 QCD+QED LO proton parton distribution set was
used~\cite{Ball:2013hta}.
Fiducial requirements were imposed using the Rivet
toolkit~\cite{Bierlich:2019rhm}.
The conditional distribution $p(\pttt, \mtt | \pthad)$ was learned via a
normalizing flow conditioned on \pthad; the cNF was trained using 200\,000
\ttbar events.

Once $p(\pttt, \mtt | \pthad)$ is approximated using the Pythia
predictions, this is applied to the $p_\mathrm{data}(\pthad)$ observed in data
to predict the joint $p(\pttt, \mtt)$ distribution by feeding samples of $\pthad \sim p_\mathrm{data}(\pthad)$ into the normalizing flow.
To achieve this, we approximate the continuous distribution of \pthad\ in data
by applying splines to the binned ATLAS measurement.
For comparison, we also apply the binned reweighting method using the binned
\pthad density from Pythia and the unfolded ATLAS data.

\begin{figure}[htpb]
    \includegraphics[width=\textwidth]{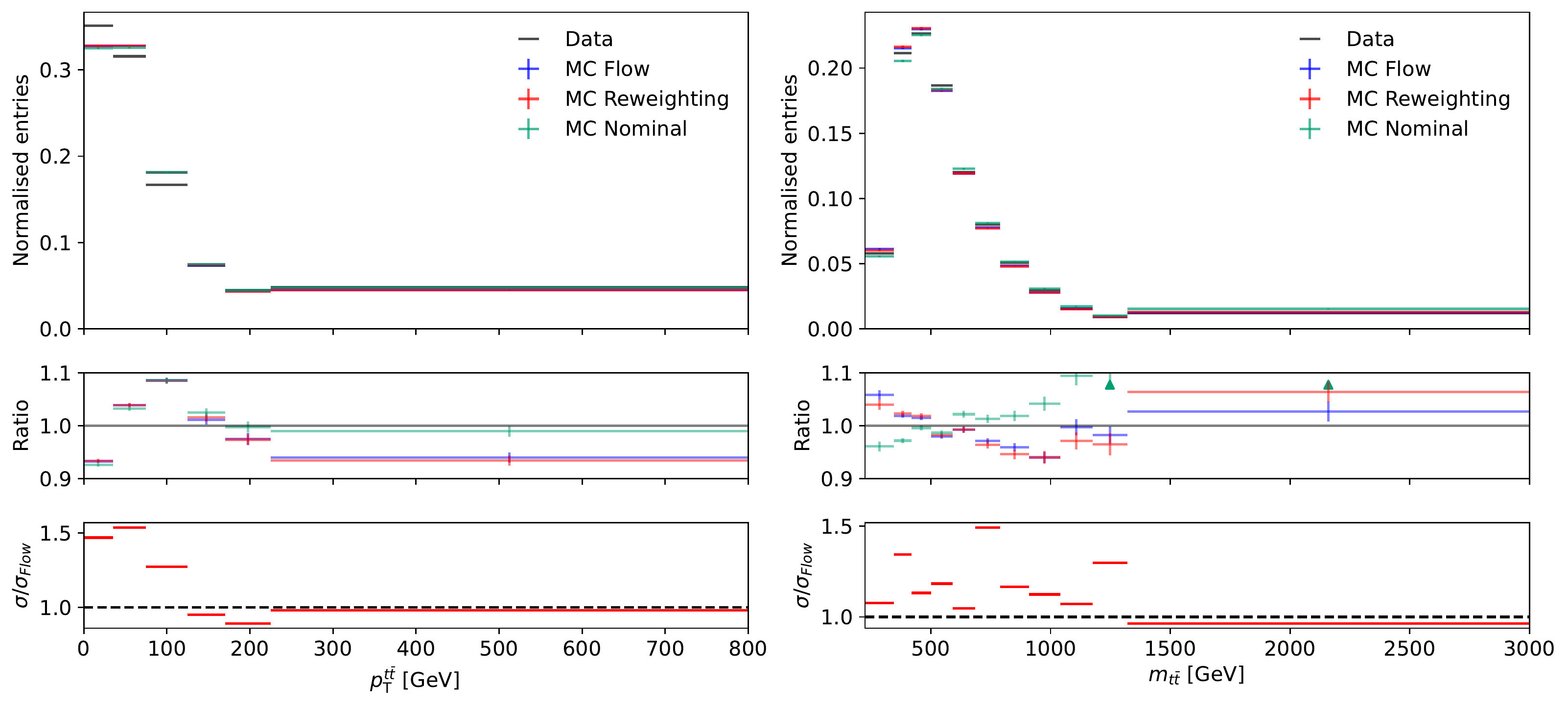}
    \caption{
        The marginals of the normalizing flow outputs compared to unfolded data.
        Statistical uncertainties are estimated using bootstraps of the training
        samples.
        The cNF approach generally achieves 25-50\% smaller
        statistical uncertainties than the binned reweighting.}
    \label{fig:top_result1}
\end{figure}

Figure~\ref{fig:top_result1} compares the outcome of the two approaches to the
unfolded data.
Neither approach faithfully reproduces the data marginals of
$p_\mathrm{data}(\pttt, \mtt)$, which we attribute to the inadequate predictive
power of the leading-order Pythia calculation.
Large differences between the resulting distributions are not expected, as in the
ideal case both the reweighting and cNF should result in the same corrected probability
distribution.
Nonetheless, correcting the \pthad\ distribution does yield a better description
of the \mtt differential cross-sections than the nominal Pythia prediction.
Crucially, our approach allows for a more precise model of the corrected Pythia, with
25-50\% smaller statistical uncertainties per bin than obtained using the
binned reweighting approach.  

\subsection{Preprocessing for neural networks}

Using this approach will be particularly useful for training other neural networks which require distributions over a subset of inputs for different classes to match, for example when using planing to reduce bias.
In comparison to reweighting and downsampling approaches, this approach does not reduce the effective training statistics, nor does it introduce a large range of weights per batch.
Additionally, it can be used to increase the effective training statistics for a neural network, which has been observed to increase performance in supervised tasks~\cite{Butter2020qhk}.

Furthermore, it can be useful when using distribution or distance based loss functions in conditional generative models which cannot easily incorporate weights.
In particular when training partially input convex neural networks~\cite{icnn}, the conditional normalizing flow can be used to guarantee the same values of non convex inputs between the input and target batch of data.
In cases where there is a difference in the distribution of the non convex inputs between the input and target data, these models can struggle to converge.

%% file: includes/conclusion.tex
In this work we have introduced an alternative approach to standard reweighting techniques, using conditional normalizing flows.
With cNFs we show it is possible to extract a model for the conditional probability distribution and generate new data from the distribution for a desired conditional distribution, removing the need to derive per event weights to modify distributions.

In comparison to binned and density estimation based reweighting techniques, this approach demonstrates better agreement to the data drawn from the true target distribution, and higher statistical precision. 
Furthermore, the statistical uncertainty on the generated data distributions come only from the number of training examples and can be calculated with bootstrapping approaches, rather than the compound of both the statistical precision on both the training and target distributions.
Due to learning the conditional distribution, using cNFs does not suffer from sparse data in high dimensions, like with binned approaches, and instead scales well to higher dimensions.
This also enables generating a much larger number of data for given conditional values, which has been seen to be beneficial with generative adversarial networks for training other neural networks.

In addition to the benefits for training other neural networks, 
we show how this approach can be applied to observed and measured distributions in high energy physics analyses. By training a cNF on the MC simulation of the signal process for the measured observables, mis-modelling in one or multiple observables compared to the observed experimental data can be accounted for and the resulting distributions can be compared to the data.
This methodology could be further expanded in reinterpretation analyses.
The distributions of non observed parameters could be changed, in order to study their impact on the agreement between recorded collision data and the theoretical predictions which depend on these parameters.
Examples for such parameters are higher order effects in the signal process, such as the modelling of top quark kinematics or monte carlo tuning parameters.
Hypothetical corrections from beyond the standard model physics that modify underlying distributions could also be studied,
such as those arising from loop effects in top quark and Higgs boson production mechanisms.

%% file: includes/appendix.tex
\addtocontents{toc}{\protect\setcounter{tocdepth}{1}}
\section{Supplementary Material}
\subsection{Detailed network architecture}
\label{app:flow_hps}

The following table show the hyperparameters of the conditional normalizing flow used in the results section

\begin{table}[htpb]
    \centering
    \begin{tabular}{cll}
    \multicolumn{1}{l}{Category} & Hyperparameter      & Value                         \\ \hline
    \multirow{5}{*}{Flow}        & Number of stacks    & 12                            \\
                                & Tails               & Linear                        \\
                                & Number of bins      & 10                            \\
                                & Tail Bound          & 3.5                             \\
                                \hline
    \multirow{4}{*}{Training}    & LR scheduler        & CosineAnnealingLR(1e-4, 1e-6) \\
                                & Number of epochs    & 500                           \\
                                & Training size       & 500.000                       \\ 
                                & Gradient clipping       & 10                       \\ 
                                \hline\hline
    \end{tabular}
    \caption{Hyperparameters of the conditional normalizing flows used for the toy examples}
    \label{tab:HP_flow_toy}
\end{table}

\begin{table}[htpb]
    \caption{Hyperparameters of the normalizing flows trained on \ttbar samples}
    \label{tab:HP_flow_ttbar}
    \centering
    \begin{tabular}{cll}
    \multicolumn{1}{l}{Category} & Hyperparameter      & Value                         \\ \hline
    \multirow{5}{*}{Flow}        & Number of stacks    & 5                            \\
                                & Tails               & Linear                        \\
                                & Number of bins      & 10                            \\
                                & Tail Bound          & 3.5                             \\
                                & Conditional Base          & True                             \\
                                \hline
    \multirow{4}{*}{Training}   & LR scheduler        & CosineAnnealingLR(1e-3, 1e-5) \\
                                & Number of epochs    & 500                           \\
                                & Training size       & 200.000                       \\ 
                                & Gradient clipping       & 10                       \\ 
                                \hline\hline
    \end{tabular}
\end{table}

\subsection{Neural network hyperparameters}
Throughout the paper we have been using neural networks as a discriminator and for the DRE (NN) method.
All the network have the same hyperparameters seen in \cref{tab:HP_NN}.

\begin{table}[htpb]
    \caption{Hyperparameters for the neural networks}
    \label{tab:HP_NN}
    \centering
    \begin{tabular}{cll}
    \multicolumn{1}{l}{Category} & Hyperparameter      & Value                         \\ \hline
    \multirow{4}{*}{Network parameters}        & Hidden layers    & 4                            \\
                                & Layer size               & 64                        \\
                                & Batch norm      & After each activation function                            \\
                                & Activation function      & Leaky ReLu                            \\
                                & Loss function          & BCE                             \\
                                \hline
    \multirow{3}{*}{Training}   & LR scheduler        & CosineAnnealingLR(1e-3, 1e-7) \\
                                & Number of epochs    & 200                           \\
                                \hline\hline
    \end{tabular}
\end{table}

\section{Additional plots}

\begin{figure}[htpb]
    \centering
    \subfloat[\centering Skewed Gaussian]{{\includegraphics[width=0.95 \textwidth]{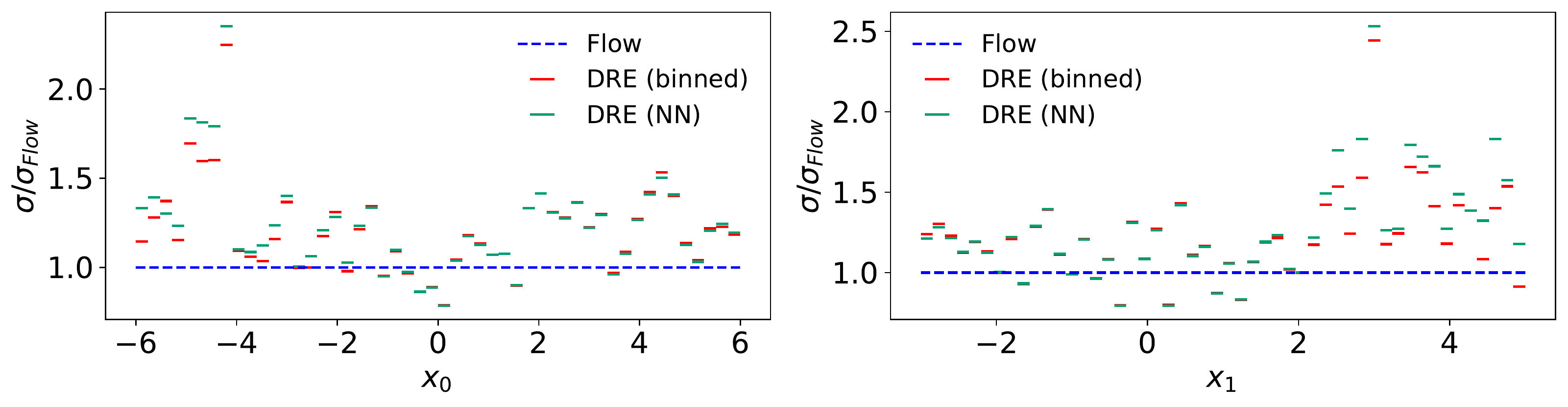} }}
    \\
    \vspace{20pt}
    \subfloat[\centering Uniform]{{\includegraphics[width=0.95 \textwidth]{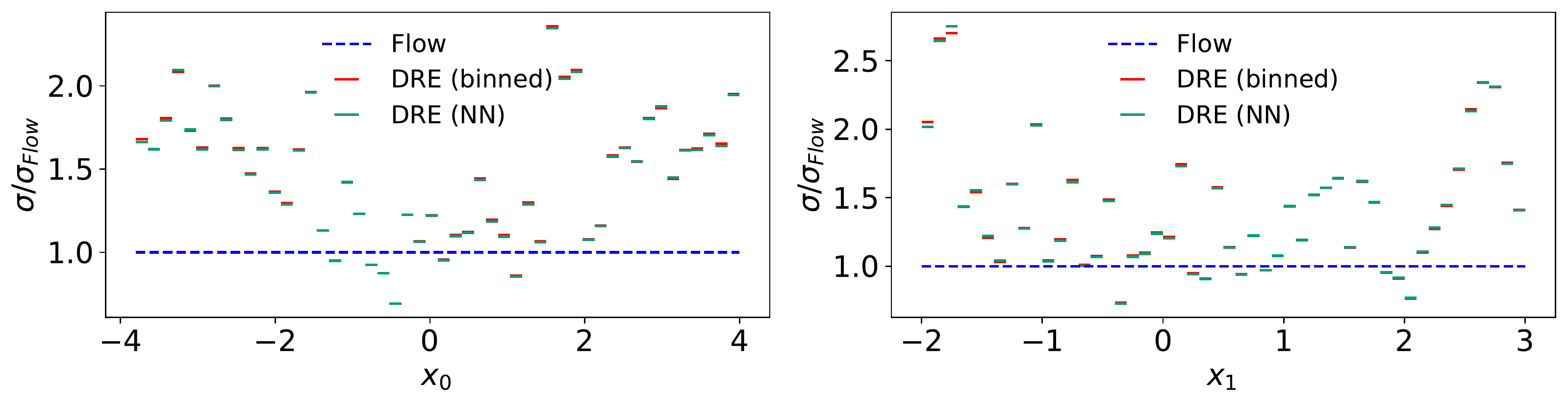} }}
    \caption{
    Comparisons are shown of estimated statistical uncertainties on binned
    marginal distributions over quantities of interest, $f_0$ and $f_1$, after
    altering the conditional quantities $c_0$ and $c_1$.
    Both the skewed Gaussian and uniform distribution examples are shown.
    The conditional normalizing flows generally result in strongly reduced
    statistical uncertainties compared to the reweight methods.
    } \label{fig:statuncerts}
\end{figure}

\begin{figure}[htpb]
    \centering
    \subfloat[\centering Skewed Gaussian]{{\includegraphics[width=0.45 \textwidth]{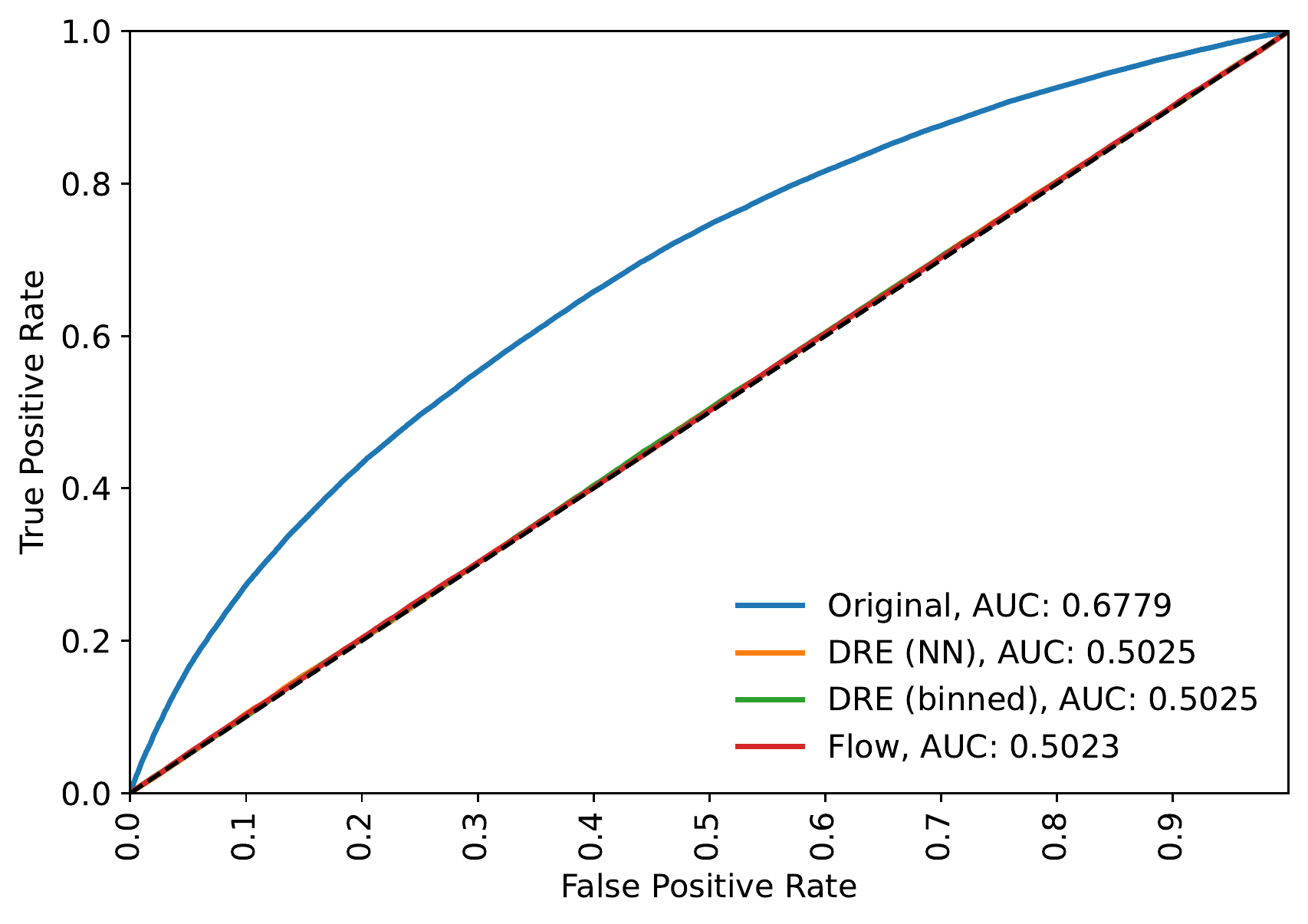} }}%
    \qquad
    \subfloat[\centering Uniform]{{\includegraphics[width=0.45 \textwidth]{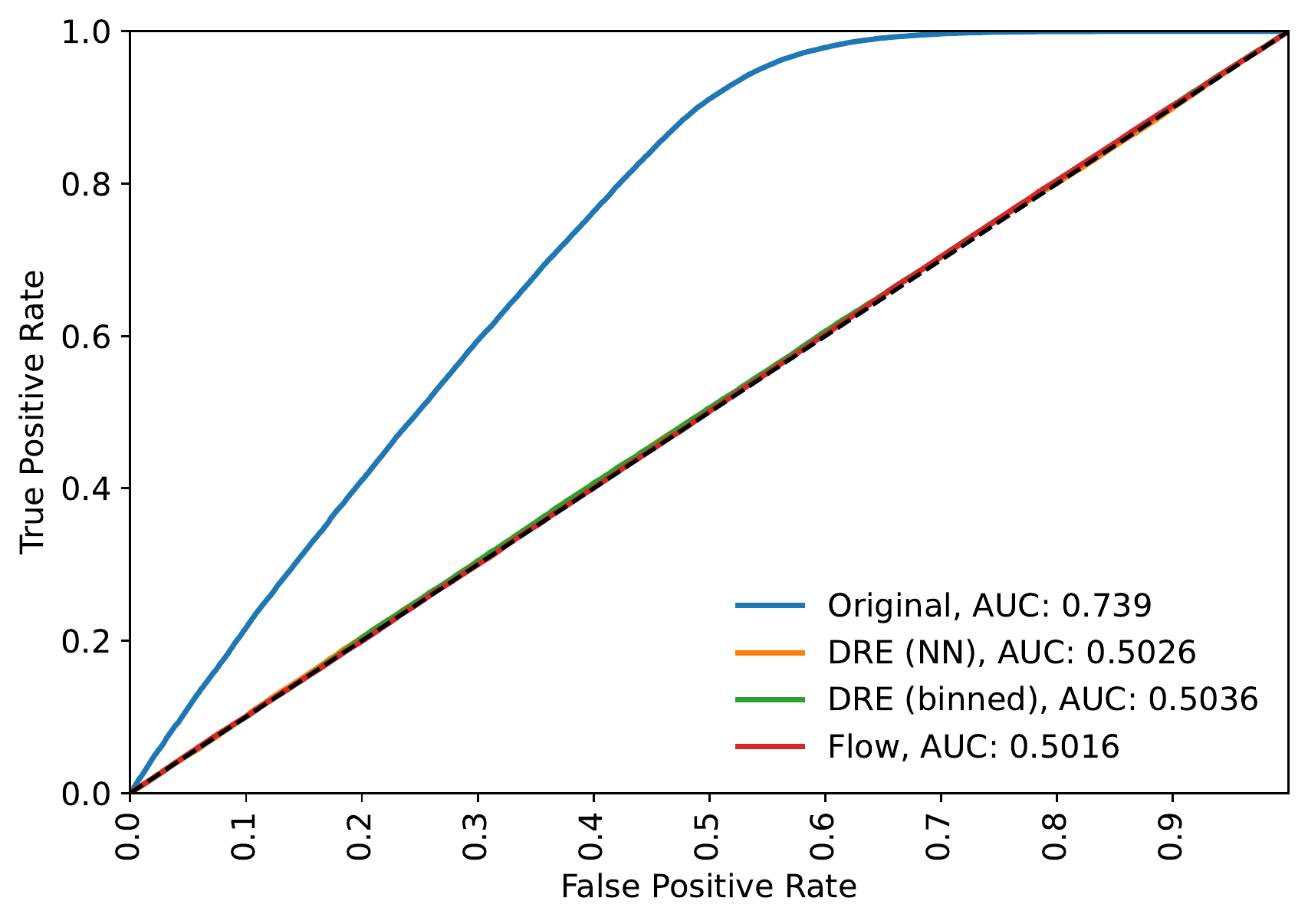} }}%
    \caption{
        Receiver operator characteristic curves for classifiers trained to discriminate between the ground
        truth distributions over $f_0$ and $f_1$, with altered distributions
        over $c_1$ and $c_0$, and those obtained through several strategies for
        correcting the conditionals' distributions.
        While a classifier can easily discriminate between the original
        distribution (produced via a standard multivariate Gaussian over $c_0$
        and $c_1$) and the altered distributions, it is unable to do so after
        correcting the conditional distributions using any of the strategies
        tested.
        } 
    \label{fig:toyrocs}
\end{figure}


